# Towards automated patient data cleaning using deep learning:

# A feasibility study on the standardization of organ labeling

Timothy Rozario, Troy Long, Mingli Chen, Weiguo Lu, and Steve Jiang

Medical Artificial Intelligence and Automation Laboratory, Department of Radiation Oncology, University of Texas Southwestern Medical Center, Dallas, TX, USA

Email: Steve.Jiang@utsouthwestern.edu

**ABSTRACT**

Data cleaning consumes about 80% of the time spent on data analysis for clinical research projects. This is a much bigger problem in the era of big data and machine learning in the field of medicine where large volumes of data are being generated. We report an initial effort towards automated patient data cleaning using deep learning: the standardization of organ labeling in radiation therapy. Organs are often labeled inconsistently at different institutions (sometimes even within the same institution) and at different time periods, which poses a problem for clinical research, especially for multi-institutional collaborative clinical research where the acquired patient data is not being used effectively. We developed a convolutional neural network (CNN) to automatically identify each organ in the CT image and then label it with the standardized nomenclature presented at AAPM Task Group 263. We tested this model on the CT images of 54 patients with prostate and 100 patients with head and neck cancer who previously received radiation therapy. The model achieved 100% accuracy in detecting organs and assigning standardized labels for the patients tested. This work shows the feasibility of using deep learning in patient data cleaning that enables standardized datasets to be generated for effective intra- and interinstitutional collaborative clinical research.

## 1. INTRODUCTION

Data cleaning consumes about 80% of the time spent on data analysis for clinical research projects (Dasu and Johnson 2003, Wickham 2014). Data cleaning improves the quality of data by detecting and removing errors and irregularities caused by inconsistencies or misspellings during data entry, missing information, and the integration of heterogeneous data sources (Rahm and Do 2000). This is not a one-time task, but one that must be performed iteratively as new issues are detected and new data are collected. Several research groups have focused on schema translation and schema integration, and some efforts have focused on identifying and eliminating duplicates (Tang, Nan 2014, Galhardas *et al* 2000b, Galhardas *et al* 2000a, Hernández and Stolfo 1998, Lee *et al* 1999, Monge 2000), but data cleaning has received very little attention (Tang 2014). This is in part due to the breadth of activities that encompass data cleaning (Wickham 2014, Raman and Hellerstein 2001, Lakshmanan *et al* 1996). Furthermore, the increasing popularity of artificial intelligence in medicine and the collection of big data (Lee and Yoon 2017, Roski *et al* 2014, Scruggs *et al* 2015) make it necessary to build automated data cleaning tools (Hu *et al* 2014, Chen and Zhang 2014, Chen *et al* 2014).

Organ labelling in radiation therapy is often inconsistent, even within an institution, due to different physician preferences, treatment planning systems, and variations that occur over time. These inconsistencies are further widened inter-institutionally. This poses a problem for clinical research, especially for multi-institutional collaborative clinical research, which relies on sharing and using standardized data across multiple sites (Cochrane *et al* 2007, Denton *et al* 2016, Gagliardi *et al* 2012, Mayo *et al* 2015, Nyholm *et al* 2016). Naming inconsistencies and the lack of naming standards in radiation therapy have led to misinterpretation of critical details, resulting in treatment errors and radiation therapy misadministration (Santanam *et al* 2013, Santanam *et al* 2012). For example, inaccurate contours or modifying or renaming structures caused miscommunication during treatment planning that resulted in 84 of the 500 (17%) misadministration events (Authority 2009). To illustrate this problem, Table 1 shows the inconsistencies in organ labeling for three patients with head and neck cancer at our institution. In patient 1, "R" and "RT" are alternatively used to represent "right" and are prefixed to the organ label, but in patient 2, "R" is placed after the organ label. Also, patient 1 appears to have two right parotids. In patients 2 and 3, numerous abbreviations appear, such as C5, T2, r1 (167), and r2 (163), among others, that were created for optimization or plan evaluation purposes. These structures do not adhere to AAPM Task Group 263 standards that require a character "z" to be prefixed to such structures so that they appear towards the end of the structure list. Also, in patients 1, 2, and 3, the right parotid gland is given three different labels: "RT PAROTID," "Parotid R," and "ParotidGland_R," respectively. This illustrates the difficulty in standardizing workflows and generating large cohorts of data for clinical research.

**Table 1.** Physician assigned organ labels for three head and neck cancer patients.

| Patient 1 | Patient 2 | Patient 3 |
|---|---|---|
| GTV-P 70 | PTV HN 66Gy | old ptv |
| GTV-N 70 | Brainstem | PTV6600_NEW |
| CTV-P 59.4 | Cochlea L | ParotidGland_L |
| R Neck lb RP 56 | Cochlea R | ParotidGland_R |
| R Parotid 56 (11) | Parotid L | BrachialPlexus_L |
| L Parotid | Parotid SUP R | BrachialPlexus_R |
| Larynx | Warm | Esophagus |
| RT PAROTID | Coverage | Spinal_cord |
| LT PAROTID | Squeeze | Brainstem |
| RT Brachial Plexus | Masseter L | normal |
| LT Brachial Plexus | Masseter R | P5940 |
| RT Cochlea | SMG L | C5 |
| LT Cochlea | SMG R | T2 |
| PTV 56 L Neck wo lb | TMJR | r66 |
| PTV 56 L Neck w lb | ICAL | r1(167) |
| RT MASSETER | R66 | r2(163) |
| LT MASSETER | NT | r3(157) |

This paper reports our work to standardize radiation therapy naming conventions of organs for specific tumor sites (prostate and head and neck) in accordance with nomenclature recommended by Mayo *et al* (2015). We used a convolutional neural network (CNN) to automatically identify each organ from the CT image before assigning labels. This is an initial effort towards automated patient data cleaning using deep learning. In the next section, we present details on the problem of organ labeling and elaborate on the CNN structure used. Then, we briefly discuss the selection of hyperparameters and the fine tuning of the model. In section 3, we present the results from prostate and head and neck cases. Finally, we give our conclusions and discuss the potential next steps for using deep learning for data cleaning in radiation oncology.

## 2. METHOD AND MATERIALS

### 2.1 Organ Labeling Problem

Organ labeling in radiation therapy is the process where the physician manually contours critical organs near the treatment site and assigns labels to them. In many institutions, there are currently no automated processes to assist physicians in this task. Our novel approach addresses the issue of organ labeling. We present a two-step process of assigning standardized labels to the organs identified: Step 1. Automatically identify the organs contoured for a specific treatment site, and Step 2. Assign standardized labels in accordance with Mayo *et al*'s standardized organ nomenclature (Mayo *et al* 2015).

In the first step, we acquire organ contour points along with their labels from the RT Structure DICOM file (Mildenberger *et al* 2002). The contour points are used as the raw input data, while the organ labels are used as the ground truth. Next, we generate 2D binary masks for each organ on a slice by slice basis and then fuse these masks onto a single image to generate a composite image that corresponds to each slice in the dataset. The extracted organ labels are used to annotate the dataset with output class labels. The set of composite images and the class labels are used as inputs to train the organ identification model. Finally, the model is tested on previously unseen composite images to determine the accuracy in identifying different organs. Formally, the organ identification model takes as input 2D composite images constructed from physician-delineated contour points where $X = [x_{1,1}, \ldots, x_{m,n}]$, such that $x_{i,j}$ corresponds to the $i$th organ of patient $j$ and outputs a sequence of predicted labels $r = [r_{i,1}, \ldots, r_{m,n}]$, such that $r_{i,j}$ can take at most one class per patient corresponding to the $i$th organ. The cross-entropy objective function (De Boer *et al* 2005, Deng 2006) is optimized for every training example to assess the model's accuracy in identifying organs in unseen test samples, as follows:

$$\mathcal{L}(X, r) = \frac{1}{n} \sum_{i=1}^{m} \sum_{j=1}^{n} \log p(R = r_{i,j} | X)$$

where $p(\cdot)$ corresponds to the probability assigned to the input $x_{i,j}$ having value $r_{i,j}$.

The organ identification problem can be modeled as a traditional multi-class classification problem that can be solved using a supervised machine learning model, such as support vector machines (SVMs) (Steinwart and Christmann 2008, Hearst *et al* 1998). Given the nature of the organ identification problem, where the input was an image and the output its corresponding organ label, a CNN was the best choice (Krizhevsky *et al* 2012). The CNN automatically extracts key localized high-dimensional features from the 2D composite images through learned weights and biases and pooling, while alleviating the effects of the curse of dimensionality (Indyk and Motwani 1998).

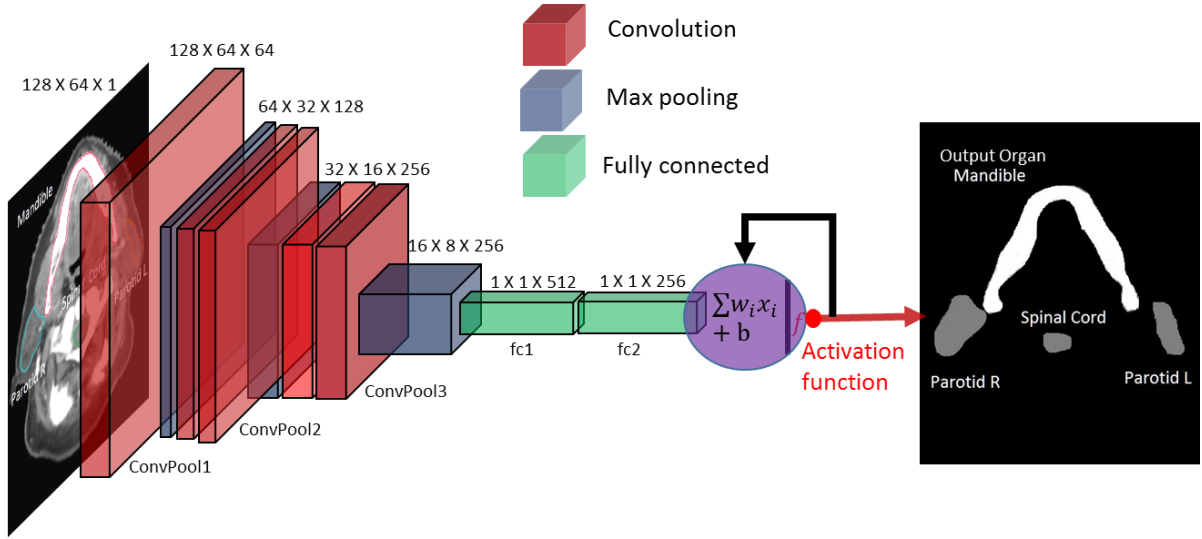

Figure 1. The CNN architecture used for organ identification in head and neck and prostate cases.

## 2.2. Patient Data

We trained and tested the organ labeling model on data from patients with head and neck or prostate cancer. For the prostate study, we collected data from 100 patients and generated a total of $\sim 40000$ composite images. We considered five critical organs for this study: prostate, bladder, rectum, left femur head, and right femur head.

**Table 2**. Details of the *CNN* architecture.

| Type | Patch Size/Stride | Output Size | Depth |
|---|---|---|---|
| Input | | 128 X 64 X 1 | 1 |
| Convolution 1 | 5 X 5/1 | 128 X 64 X 64 | 1 |
| Max Pool 1 | 5 X 5/2 | 64 X 32 X 64 | 0 |
| Convolution 2 | 5 X 5/1 | 64 X 32 X 128 | 1 |
| Max Pool 2 | 5 X 5/2 | 32 X 16 X 128 | 0 |
| Convolution 3 | 5 X 5/1 | 32 X 16 X 256 | 1 |
| Max Pool 3 | 5 X 5/2 | 16 X 8 X 256 | 0 |
| Fully-connected 1 | | 1 X 1 X 512 | 1 |
| Fully-connected 2 | | 1 X 1 X 256 | 1 |
| Activation Function | Softmax | | |
| Optimizer | Adam | Default parameters | |

**Table 3**. Summary of hyperparameters used.

| Hyperparameter | Value |
|---|---|
| Batch Size | 200 |
| Epochs | 10 |
| Learning Rate | $10e^{-3}$ |
| Regularization | $10e^{-3}$ |
| Momentum | 0.99 |
| Decay | 0.99 |

For the head and neck study, we collected data from 54 patients and generated a total of ∼ 45000 composite images. We considered nine critical organs for this study: spinal cord, brainstem, larynx, mandible, esophagus, left parotid, right parotid, left cochlea, and right cochlea.

The 2D composite images used as input was a novel representation of the raw data acquired in the form of organ contour points. Each composite image consisted of organ masks that had contour points for the specific slice in consideration. Finally, we changed the pixel intensities (pixel manipulation) for each composite image, such that, for a specific organ class, that organ mask was assigned the highest pixel value, while the remaining organ masks present in that composite image were assigned a pixel value of half that of the organ class. This enabled the model to learn high-dimensional spatial, structural, and intensity features. Figure 3(a) shows an example of a constructed composite image with pixel manipulation for the organ class Mandible. In this image, the Mandible has a high pixel intensity of 1, while the Parotid L/R and Spinal Cord have pixel values of 0.5.

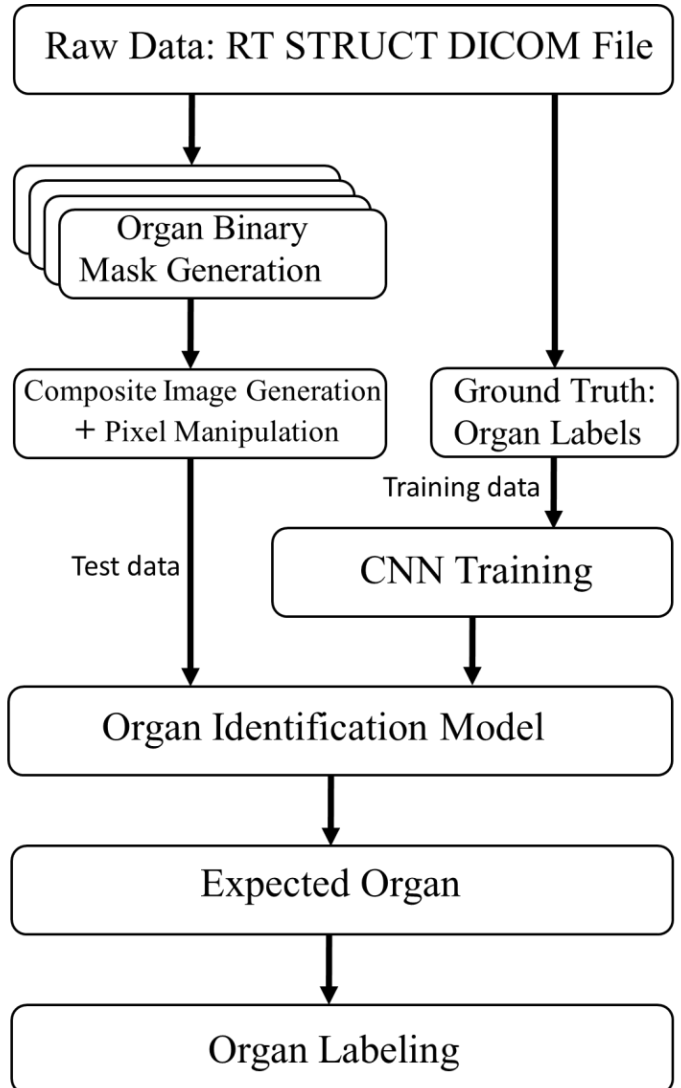

Figure 2. An overview of the different steps involved in identifying and labeling organs for patients with head and neck or prostate cancer.

### 2.3 Model Architecture

The CNN model used in the organ identification step consists of three convolutional and subsampling layers and two fully connected layers (Fig. 2). Neurons constitute the convolutional/fully-connected layers that have learnable weights and biases. The input to the CNN was an $n \times m \times r$ composite image, where $n$ was the height, $m$ was the width, and $r$ was the number channels, which was set to 1 (grayscale images). The convolutional layer had $k$ kernels of size $j \times j \times k$, where $j$ was set to 3, and $k > r$. The value of $k$ was doubled at each of the three convolutional layers, such that $k = 64, 128, 256$ at layers 1, 2, and 3, respectively. The filter size enabled the extraction of locally connected structures that were convolved with the image to produce feature maps. These feature maps were then subsampled using $2 \times 2$ contiguous max pooling to retain the most prominent localized features in the image. The resulting feature maps decreased in size by half after the completion of each of the three max pooling layers ($128 \times 64$ to $64 \times 32$; $64 \times 32$ to $32 \times 16$; $32 \times 16$ to $16 \times 8$). Prior to max pooling, an additive bias and nonlinearity was applied to each feature map. We used the rectified linear activation function (ReLU) in the hidden layers, which is piece-wise linear and saturates at zero when the input $z$ is less than 0. Finally, we used two fully connected layers along with the softmax activation function in the output layer to provide the probabilities of the computed output classes (Dunne and Campbell 1997).

The composite images generated enabled the CNN to effectively learn high-dimensional localized features and leveraged its ability to extract key image-contrast features because of the pixel intensity manipulation performed on the composite images. Creating composite images for each organ class increased the size of the dataset by a factor of 9 (the number of organs considered) for head and neck cases and by a factor of 5 for prostate cases. For this study, the data samples were split into training and testing sets that were randomly selected with no overlapping and no bias among the organ classes. A data split ratio of 80/20 was used for training and testing, respectively. For both prostate and head and neck cases, the physician-annotated contours from the CT simulations were used as the ground truth. The model reported multi-class $9D$ and $5D$ vectors of probabilities that corresponded to the organs considered for head and neck and prostate cases, respectively.

### 2.4 Model Parameter Optimization

We performed network optimization strategies, such as batch normalization, after each convolutional layer and before applying the ReLU activation function (Ioffe and Szegedy 2015). We used a sparse implementation of the Adam algorithm as the optimizer with default parameters (Kingma and Ba 2014) and a random search approach to identify the best hyperparameter values (Bergstra and Bengio 2012). This not only improved the accuracy of organ identification, but it also reduced the number of batch iterations required to reach this accuracy. See Table 3 for details on the hyperparameter values.

Once the CNN model has identified the organ, the next step is to assign a standardized label to it. For this, we created site-specific templates of standardized names for the critical organs, as recommended by Mayo *et al* (2015). Next, we mapped the output organ classes to the corresponding names in the template to assign a standardized label for each of the organs detected. See Figure 2 for an overview of the steps involved in identifying the organs and labeling them.

## 3. RESULTS

For this study, we considered patients with head and neck and prostate cancer. For the prostate cases, the organ labeling model learned high-dimensional features that identified the 5 organ classes with 100% accuracy and assigned standardized labels as recommended by Mayo *et al* (2015). Despite inconsistencies

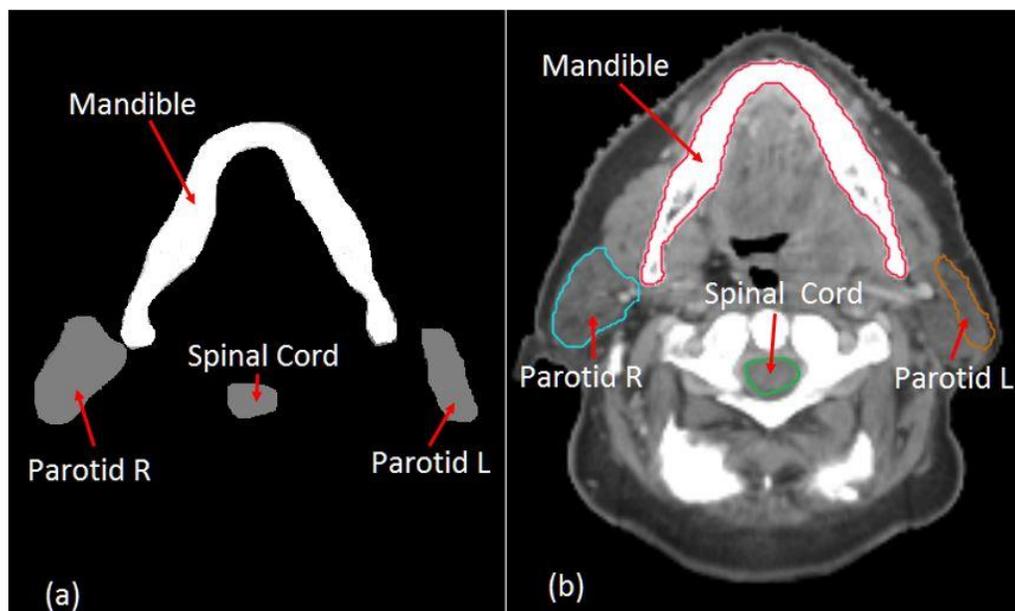

Figure 3. (a) represents the composite image from a patient with head and neck cancer for Mandible with pixel intensity 1.0 while other organ masks have pixel intensity 0.5, (b) represents the corresponding CT slice for the same patient with the organs contoured and labeled.

n organ contouring and labeling, the model extracted key features that allowed for successful organ identification. The CNN was trained on batches of 150 data samples after normalization. The data was

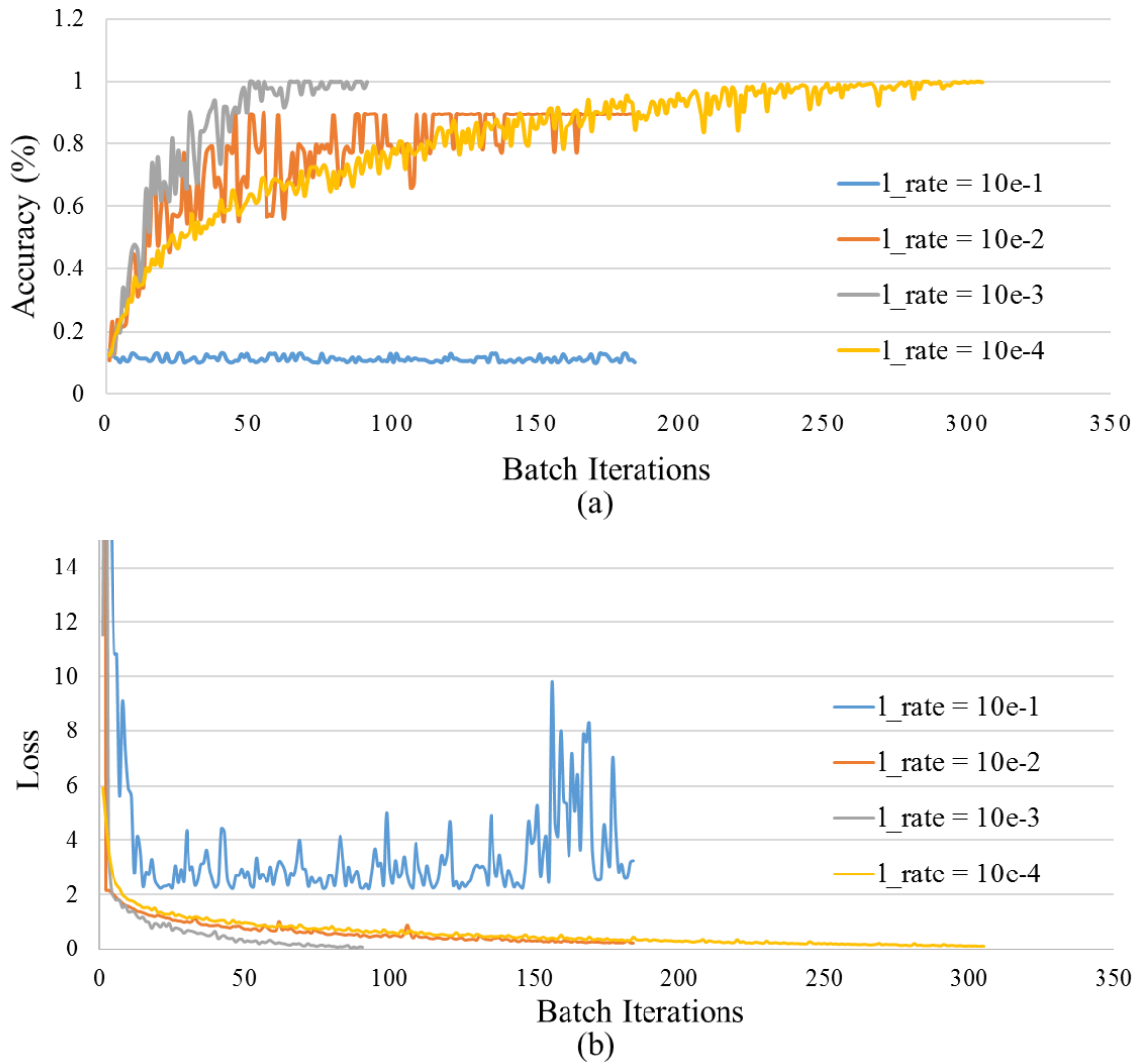

(a)

(b)

Figure 4 (a) shows the accuracy that our model reported in identifying the organs for head and neck cases for different learning rates, and (b) shows the cross-entropy loss achieved by our model for the four learning rates.

randomly split into training data (80%) and testing data (20%) with no overlap and no data bias. The CNN model reported an accuracy of 100% within 5 epochs.

For the head and neck cases, the organ labeling model was able to learn high-dimensional features that identified the 9 organ classes with 100% accuracy and assigned standardized labels as recommended by Mayo *et al* (2015). Despite inconsistencies in organ contouring and labeling, the model extracted key features that allowed for successful organ identification. The CNN was trained on batches of 150 data samples after normalization. The data was randomly split into training data (80%) and testing data (20%) with no overlap and no data bias. The CNN model reported an accuracy of 100% within 10 epochs.

The number of slices considered for each patient was determined independently. First, we computed the minimum and maximum range of CT slice numbers for each organ class. Then, we determined the biggest range of CT slice numbers that contained contour points from all the organ classes. The number of slices per patient was equal to the number of composite images generated. Based on this methodology, the slices having contour points from multiple organs resulted in composite images with multiple organ masks. In some cases, the composite image contained a single organ mask, which was considered the most difficult case. Yet, the CNN model identified the organ accurately with the help of high-dimensional intensity features generated by the pixel intensity manipulation performed on the composite images. Hence, a combination of structural, spatial, and intensity features enabled the model to learn inter-image relationships, despite the input being 2D composite images.

The model's accuracy and training speed changed drastically when using different learning rates. Figure 4(a) illustrates the accuracy in identifying the different organs, and 4(b) shows the loss achieved for different learning rates. For a learning rate of $10e^{-1}$, the accuracy was flat at 10%, which represented a failure to identify the organs. For a learning rate of $10e^{-2}$, the model reported an accuracy of 89% then plateaued at the same level. The model took 184 batch iterations to converge. For a learning rate of $10e^{-4}$, the *CNN* reported an accuracy of 100%, but it took 305 batch iterations to converge. For a learning rate of $10e^{-3}$, the CNN performed very efficiently, taking 90 batch iterations to converge with a 100% organ identification accuracy. Thus, $10e^{-3}$ was the most effective learning rate for training the model.

We saw similar trends in the cost minimization for the different learning rates. With a learning rate of $10e^{-1}$, the cost remained flat at 3.52 after running the model for 184 batch iterations. This learning rate did not minimize the distance substantially between the identified organs and corresponding ground truth class labels. The cost function reduced the distance to 0.234 in 184 batch iterations with a learning rate of $10e^{-2}$, while a cost reduction of 0.129 was reported by the model at a learning rate of $10e^{-4}$. The best

cost reduction of 0.092 was seen with a learning rate of $10e^{-3}$, which also converged the quickest, in 90

batch iterations.

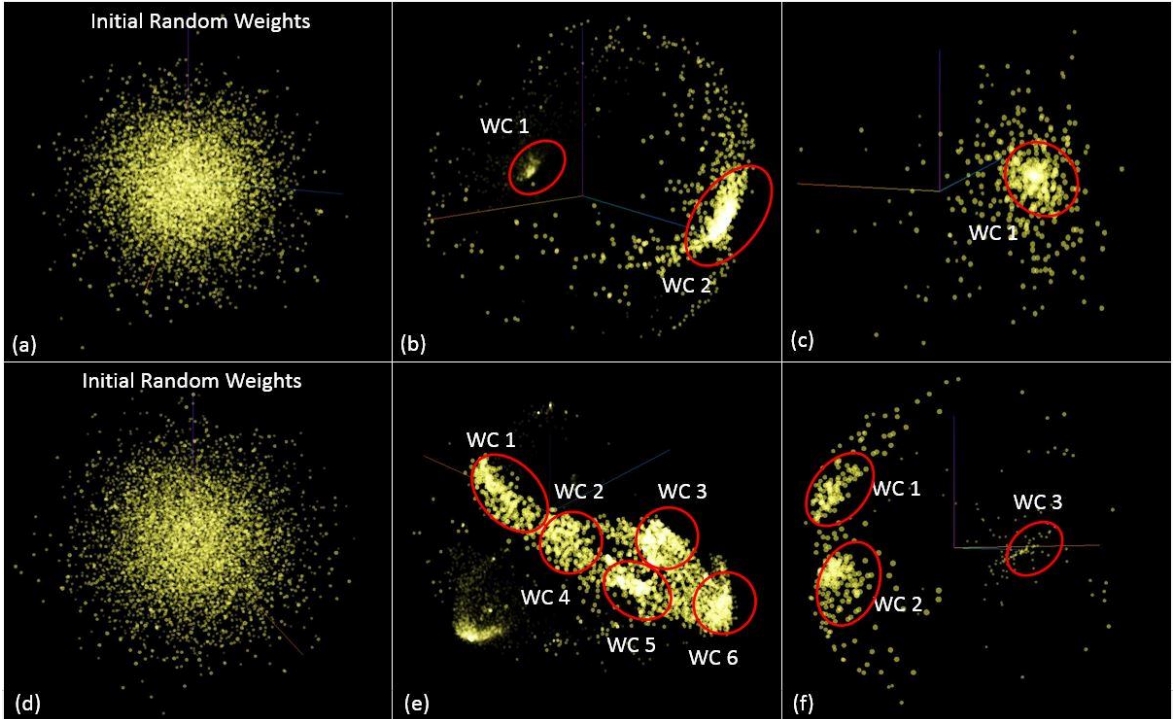

Figure 5. Presents a visualization of the CNN model weight clusters (WC) for two learning rates (top row -
$10e^{-3}$, bottom row -$10e^{-1}$ ) in 3D space for the top three PCA modes. (a) and (d) show initial random weights
for the two runs, (b) and (e) show the paritally learned weights after the completion of the convolutional step, and
(c) and (f) show the final set of weights that are computed after the fully-connected layers have finished learning.

To understand how the CNN learned high-dimensional features, we visually illustrated the weights of the model at different stages of the training process for two learning rates: $10e^{-3}$ and $10e^{-1}$ (Fig. 5). We

displayed the initial randomized weights and the learned weights after the convolutional step and the fully-

connected network optimization in 3D space after computing the top 3 principal component analysis (PCA)

modes (Abdi and Williams 2010, Wold *et al* 1987) using Tensorflow's Tensorboard (Abadi *et al* 2016).

This helped us understand how the model was performing and how to better fine tune it. The top row shows how the weights were optimized while training the CNN model for the learning rate $10e^{-3}$. We observed two weight clusters (WC) after the convolutional step and a single WC after the fully-connected step. This indicates that the model extracted key localized features and accurately identified organ classes. During the computation of the top 3 PCA modes for learning rate $10e^{-3}$, we observed 4% total variance for the initial random weights and 98% total variance after the convolutional step, while a total variance of 99.9% was observed after the completion of the fully-connected network optimization. This indicated that the model had learned high-dimensional features. In Figure 5, the bottom row illustrates how the weights were optimized while training the model for a learning rate of $10e^{-1}$. Here, we noticed many WCs after the convolutional step, indicating that the model was learning conflicting features. Our observation of WCs after the fully-connected step emphasized this further. This coincided with the results observed from Figure 4, where the model failed to converge and identified organs with an accuracy of roughly 10%.

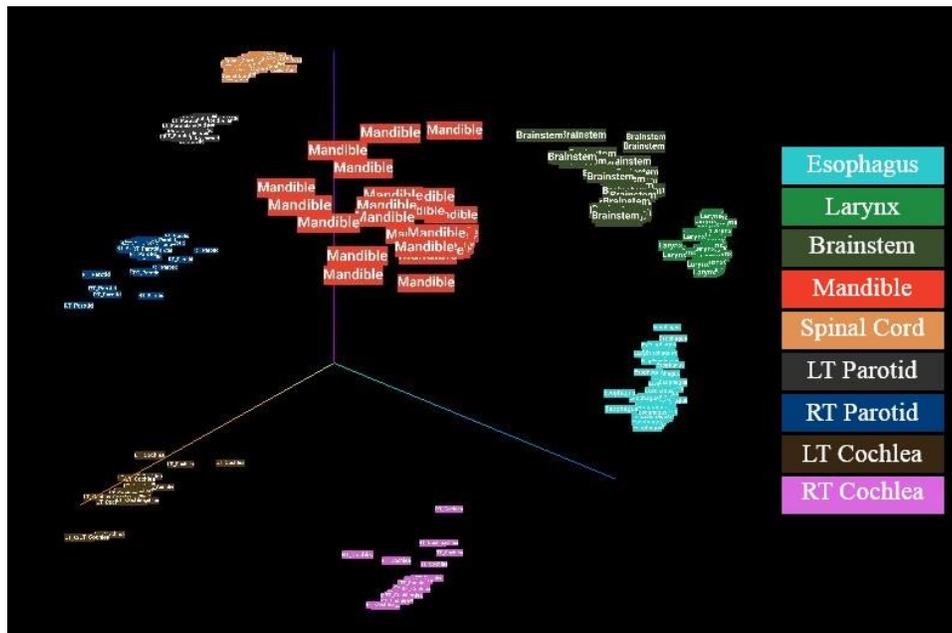

Figure 6. Presents a 3D visual representation of the organs identified in patients with head and neck cancer. The nine organs are color-coded to illustrates the clusters they form when the top three PCA modes are computed.

In Figure 6, we visualize previously unseen test data samples that have been identified by the CNN model. The organ classes are represented in 3D space with color coding assigned to each organ class after computing the top three PCA modes. The total variance of the top 3 PCA modes was 93.8%, while the individual variances were 74.6%, 17.5%, and 1.7%, respectively. The nine organ classes were well-clustered and separated from each other. This representation of the test data provides a way of visualizing not only the variations that occur between class clusters, but also the subtle differences that occur within a class cluster.

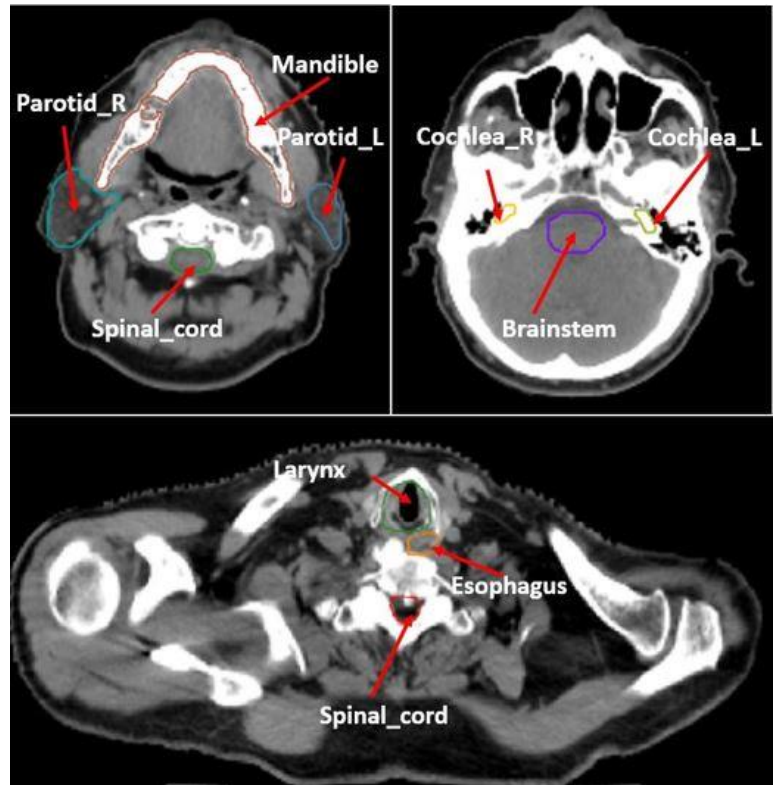

Figure 7. Shows the CT representation of the organs identified (contoured) and labeled on numerous slices for a patient with head and neck cancer.

Figure 7 provides a CT slice visualization of the organs that have been accurately identified and labeled by the CNN model on patients with head and neck cancer.

## 4. DISCUSSION AND CONCLUSIONS

Traditionally, patient data cleaning has accounted for approximately 80% of the time spent on data analysis in clinical research projects. In this work, we reported an initial effort towards automated patient data cleaning, specifically, the task of standardizing organ labels. Assigning standardized labels to the organs is crucial to avoiding the misinterpretation of critical details, which can result in treatment errors and radiation therapy misadministration. We have shown that deep learning can achieve accurate organ identification and standardized labeling without manual intervention, thus eliminating errors and inconsistencies in labeling patient data. We tested the model on patients with head and neck or prostate cancer. In total, 154 patients' data (100 prostate cases and 54 head and neck cases) were utilized in this study. Our model identified organ contours with an accuracy of 100% and assigned standardized labels.

In our future work on patient data cleaning, we focus on two areas: increasing the number of organs labeled for head and neck cases and contouring errors. We intend to incorporate a comprehensive list of head and neck organs (this was not possible in this study due to the small patient data set that we had to work with) as prescribed by our institution. Our goal is to develop a 3D model that accepts volumetric anatomical data to identify organs automatically. This will allow the model to learn high-dimensional volumetric features. We intend to add this model to our current clinical system to automate the procedure of standardizing organ labels and retrospectively applying the model to existing patient data to promote clinical research, especially multi-institutional collaborative clinical research.

We are also working to identify and correct contour errors using deep learning. Errors such as incorrect contouring, contour-label mismatch, and volume fill (not closing the manually contoured volume), among others, could result in treatment errors and significantly lengthen the treatment planning process. Currently, at our institution, it takes 7-10 days to plan treatment for head and neck cases after the CT simulation. If contouring errors are detected, the required rework would increase treatment planning time between 28% to 57%. In addition, this also adds logistic complexity to the whole process, because it involves many people, including the physician, dosimetrist, physicists, and the Quality Assurance (QA) team, hence the need for automated tools.

**Acknowledgements**

We would like to thank Drs. Mu-Han Lin, Dan Nguyen, and Jonathan Feinberg for helpful discussions and proof reading of the manuscript.

# REFERENCES

Abadi, M, Agarwal, A, Barham, P, Brevdo, E, Chen, Z, Citro, C, Corrado, G S, Davis, A, Dean, J and Devin, M 2016 Tensorflow: Large-scale machine learning on heterogeneous distributed systems *arXiv preprint arXiv:1603.04467*

Abdi, H and Williams, L J 2010 Principal component analysis *Wiley interdisciplinary reviews: computational statistics* **2** 4 433-59

Authority, P P S 2009 Errors in radiation therapy *Pennsylvania Patient Safety Advisory* **6** 3 87-92

Bergstra, J and Bengio, Y 2012 Random search for hyper-parameter optimization *Journal of Machine Learning Research* **13** Feb 281-305

Chen, C P and Zhang, C 2014 Data-intensive applications, challenges, techniques and technologies: A survey on Big Data *Inf. Sci.* **275** 314-47

Chen, M, Mao, S and Liu, Y 2014 Big data: A survey *Mobile Networks and Applications* **19** 2 171-209

Cochrane, L J, Olson, C A, Murray, S, Dupuis, M, Tooman, T and Hayes, S 2007 Gaps between knowing and doing: understanding and assessing the barriers to optimal health care *J. Contin. Educ. Health Prof.* **27** 2 94-102

Dasu, Tand Johnson, T 2003 *Exploratory Data Mining and Data Cleaning* : John Wiley & Sons)

De Boer, P, Kroese, D P, Mannor, S and Rubinstein, R Y 2005 A tutorial on the cross-entropy method *Annals of operations research* **134** 1 19-67

Deng, L 2006 *The cross-entropy method: a unified approach to combinatorial optimization, Monte-Carlo simulation, and machine learning*

Denton, T R, Shields, L B, Hahl, M, Maudlin, C, Bassett, M and Spalding, A C 2016 Guidelines for treatment naming in radiation oncology *Journal of applied clinical medical physics* **17** 2 123-38

Dunne, R Aand Campbell, N A 1997 On the pairing of the softmax activation and cross-entropy penalty functions and the derivation of the softmax activation function *Proc. 8th Aust. Conf. on the Neural Networks, Melbourne, 181*

Gagliardi, A R, Brouwers, M C and Bhattacharyya, O K 2012 The guideline implementability research and application network (GIRAnet): an international collaborative to support knowledge exchange: study protocol *Implementation Science* **7** 1 26

Galhardas, H, Florescu, D, Shasha, D and Simon, E 2000a AJAX: an extensible data cleaning tool *ACM Sigmod Record* : ACM) pp 590

Galhardas, H, Florescu, D, Shasha, D and Simon, E 2000b Declaratively cleaning your data using AJAX *In Journees Bases De Donnees* : Citeseer)

Hearst, M A, Dumais, S T, Osuna, E, Platt, J and Scholkopf, B 1998 Support vector machines *IEEE Intelligent Systems and their applications* **13** 4 18-28

Hernández, M A and Stolfo, S J 1998 Real-world data is dirty: Data cleansing and the merge/purge problem *Data mining and knowledge discovery* **2** 1 9-37

Hu, H, Wen, Y, Chua, T and Li, X 2014 Toward scalable systems for big data analytics: A technology tutorial *IEEE access* **2** 652-87

Indyk, Pand Motwani, R 1998 Approximate nearest neighbors: towards removing the curse of dimensionality *Proceedings of the Thirtieth Annual ACM Symposium on Theory of Computing* : ACM) pp 604-13

Ioffe, Sand Szegedy, C 2015 Batch normalization: Accelerating deep network training by reducing internal covariate shift *International Conference on Machine Learning* pp 448-56

Kingma, D and Ba, J 2014 Adam: A method for stochastic optimization *arXiv preprint arXiv:1412.6980*

Krizhevsky, A, Sutskever, I and Hinton, G E 2012 Imagenet classification with deep convolutional neural networks *Advances in Neural Information Processing Systems* pp 1097-105

Lakshmanan, L V, Sadri, F and Subramanian, I N 1996 SchemaSQL-a language for interoperability in relational multi-database systems *Vldb* pp 239-50

Lee, C H and Yoon, H J 2017 Medical big data: promise and challenges *Kidney Res. Clin. Pract.* **36** 1 3-11

Lee, M, Lu, H, Ling, T W and Ko, Y T 1999 Cleansing data for mining and warehousing *Dexa* : Springer) pp 751-60

Mayo, C, Moran, J, Xiao, Y, Bosch, W, Matuszak, M, Marks, L, Miller, R, Wu, Q, Yock, T and Popple, R 2015 AAPM Task Group 263: tackling standardization of nomenclature for radiation therapy *International Journal of Radiation Oncology• Biology• Physics* **93** 3 E383-4

Mildenberger, P, Eichelberg, M and Martin, E 2002 Introduction to the DICOM standard *Eur. Radiol.* **12** 4 920-7

Monge, A E 2000 Matching algorithms within a duplicate detection system *IEEE Data Eng.Bull.* **23** 4 14-20

Nyholm, T, Olsson, C, Agrup, M, Björk, P, Björk-Eriksson, T, Gagliardi, G, Grinaker, H, Gunnlaugsson, A, Gustafsson, A and Gustafsson, M 2016 A national approach for automated collection of standardized and population-based radiation therapy data in Sweden *Radiotherapy and Oncology* **119** 2 344-50

Rahm, E and Do, H H 2000 Data cleaning: Problems and current approaches *IEEE Data Eng.Bull.* **23** 4 3-13

Raman, V and Hellerstein, J M 2001 Potter's wheel: An interactive data cleaning system *Vldb* pp 381-90

Roski, J, Bo-Linn, G W and Andrews, T A 2014 Creating value in health care through big data: opportunities and policy implications *Health. Aff. (Millwood)* **33** 7 1115-22

Santanam, L, Brame, R S, Lindsey, A, Dewees, T, Danieley, J, Labrash, J, Parikh, P, Bradley, J, Zoberi, I and Michalski, J 2013 Eliminating inconsistencies in simulation and treatment planning orders in radiation therapy *International Journal of Radiation Oncology\* Biology\* Physics* **85** 2 484-91

Santanam, L, Hurkmans, C, Mutic, S, van Vliet-Vroegindeweij, C, Brame, S, Straube, W, Galvin, J, Tripuraneni, P, Michalski, J and Bosch, W 2012 Standardizing naming conventions in radiation oncology *International Journal of Radiation Oncology\* Biology\* Physics* **83** 4 1344-9

Scruggs, S B, Watson, K, Su, A I, Hermjakob, H, Yates, J R,3rd, Lindsey, M L and Ping, P 2015 Harnessing the heart of big data *Circ. Res.* **116** 7 1115-9

Steinwart, I and Christmann, A 2008 *Support Vector Machines* : Springer Science & Business Media)

*Tang, N* 2014 Big data cleaning *Asia-Pacific Web Conference* : Springer) pp 13-24

Wickham, H 2014 Tidy data *Journal of Statistical Software* **59** 10 1-23

Wold, S, Esbensen, K and Geladi, P 1987 Principal component analysis *Chemometrics Intelig. Lab. Syst.* **2** 1-3 37-52